\documentclass{aa}

\usepackage[]{aalongtable}

\begin{document}

\input{psfig}


\title{VLT-UVES abundance analysis of four  giants  in  NGC 6553\thanks{Observations 
collected both  at the European  Southern  Observatory,  Paranal, and La 
Silla,  Chile  (ESO programmes 65.L-0340, 65.L-0371, 67.D-0489, and 69.D-0582)} }

%
\author{
A. Alves-Brito\inst{1}
\and
B. Barbuy\inst{1}
\and
M. Zoccali\inst{2}
\and
D. Minniti\inst{2}
\and
S. Ortolani\inst{3}
\and
V. Hill\inst{4}
\and
A. Renzini\inst{5}
\and
L. Pasquini\inst{6}
\and
E. Bica\inst{7}
\and
R.M. Rich\inst{8}
\and
J. Mel\'endez\inst{9}
\and
Y. Momany\inst{3}
}
\offprints{A. Alves-Brito}
\institute{
Universidade de S\~ao Paulo, IAG, Rua do Mat\~ao 1226,
Cidade Universit\'aria, S\~ao Paulo 05508-900, Brazil\\
 e-mail: abrito@astro.iag.usp.br, barbuy@astro.iag.usp.br
\and
Universidad Catolica de Chile, Department of Astronomy \& Astrophysics,
Casilla 306, Santiago 22, Chile\\
e-mail: dante@astro.puc.cl, mzoccali@astro.puc.cl
\and
Universit\`a di Padova, Dipartimento di Astronomia, Vicolo dell'Osservatorio
2, I-35122 Padova, Italy\\
 e-mail: ortolani@pd.astro.it
\and
Osservatorio Astronomico di Padova, Vicolo dell'Osservatorio 5, I-35122 Padova,  Italy; e-mail:
arenzini@pd.astro.it
\and 
 European Southern Observatory, Karl Schwarzschild Strasse 2, 85748
Garching bei M\"unchen, Germany \\
e-mail:  lpasquin@eso.org
\and 
 European Southern Observatory, Karl Schwarzschild Strasse 2, 85748
Garching bei M\"unchen, Germany \\
e-mail:  arenzini@eso.org, lpasquin@eso.org
\and
Universidade Federal do Rio Grande do Sul, Departamento de Astronomia, 
CP 15051, Porto Alegre 91501-970, Brazil; e-mail: bica@if.ufrgs.br
\and
UCLA, Department of Physics \& Astronomy, 8979 Math-Sciences Building,
Los Angeles, CA 90095-1562, USA;\\
e-mail: rmr@astro.ucla.edu
\and
Australian National University, Australia; e-mail: jorge@mso.anu.edu.au
}

\date{Received ; accepted }

 
  \abstract
{Metal-rich globular clusters trace the formation of bulges.
Abundance ratios in the metal-rich globular clusters such as NGC 6553
 can constrain the formation timescale of the Galactic bulge.
}
   { The  purpose  of this  study is determine the metallicity and elemental  ratios in individual
stars of the metal-rich bulge globular cluster NGC 6553. }
{A detailed abundance analysis of four giants in 
 NGC 6553 is carried out, based on optical high-resolution
\'echelle spectra obtained with UVES at the ESO VLT-UT2 Kueyen telescope.
}
{
A metallicity [Fe/H]= $-$0.20 dex is derived, together with $\alpha$-element
enhancement of Mg and Si ([Mg/Fe]=+0.28, [Si/Fe]=+0.21), 
solar Ca and Ti ([Ca/Fe]=+0.05, [Ti/Fe]=$-$0.01), and a mild enhancement of the r-process element Eu 
with [Eu/Fe] = $+$0.10. A mean heliocentric radial velocity of $-$1.86 km
s$^{-1}$ is measured. We compare our results with previous investigations of
the cluster.
}
   {}

\maketitle


\section{Introduction} 

Metal-rich red globular clusters trace the build up of bulges and
possibly of disks (Brodie \& Strader 2006).
They may be formed  during an early collapse
 (Forbes et al. 1997; C\^ot\'e et al. 1998); otherwise,
in hierarchical scenarios the metal-rich globular clusters
would be direct
probes of the merging events and assembly of the host galaxy (Ashman
\& Zepf 1992; Beasley et al. 2002; Bekki 2005).
Another possibility is bulge formation via the secular
evolution of a bar (Kormendy \& Kennicutt 2004).
 Their detailed study
should help constrain models of globular cluster formation in
the early Galaxy.

Several studies have tried to infer some of the main
properties of NGC 6553.
Minniti (1995) argues that the metal-rich globular clusters in the inner
spheroid are associated with the galactic bulge.  
Zoccali et al. (2001, 2003) determined the proper motion of NGC 6553 and
conclude that this cluster follows the mean rotation of both disk and bulge
stars, whereas Dinescu et al. (2003) 
conclude that this cluster belongs to a rotationally
supported disk system. 
Ortolani et al. (1995) show that NGC 6528 and NGC 6553 are old.
Another tool for better understanding the origin of
these metal-rich clusters is the chemical composition of individual member
stars.

The globular clusters in the bulge  NGC 6553, NGC 6528, NGC 6440,
and Liller 1 are among the most metal-rich clusters in the Galaxy
 (Barbuy et al. 1998).
Ortolani et al. (1990) point out
the turnover and fainter red giant branch (RGB) of NGC 6553, relative to
less metal-rich clusters, indicating that TiO blanketing should contribute
to the RGB behaviour, being in itself an indicator of high  metallicity.
These metal-rich clusters constitute the only available
calibrators for metallicity and abundance ratio measurement
methods based on low-resolution spectra and/or population synthesis
of composite spectra of star clusters and galaxies.

NGC 6553 is the most studied bulge globular cluster, and yet only
few stars were analysed at high resolution.
The CCD analyses were carried out for one giant by
Barbuy et al. (1992) and two giants by Barbuy et al. (1999) 
at moderate resolution
(R$\sim$20 000),  five red horizontal
branch stars by Cohen et al. (1999) at high resolution (R$\sim$34 000), 
whereas in the H band two giants were analysed by
Origlia et al. (2002) at moderate resolution (R$\sim$25 000)
 and five giants by
Mel\'endez et al. (2003) at high resolution (R$\sim$50 000).
Carretta et al. (2001) revised the Cohen et al. (1999) analysis.
These previous results gave metallicities in the range
 $-$0.55 $<$ [Fe/H] $<$ $-$0.06.

We present the analysis of four giants 
 based on high-resolution (R = 55 000) spectra obtained at the 
VLT-UT2 Kueyen telescope equipped with the UVES spectrograph.
These are the highest resolution observations available for stars in
this cluster. Since we are dealing with cool metal-rich giants,
high-resolution spectroscopy is mandatory.

The observations and radial velocity derivation are described in Sect. 2. In
Sect. 3 the stellar parameters are presented, the abundances are
derived in Sect. 4, the results are discussed in Sect. 5, and a summary
is given in Sect. 6.
\section{Observations} 
\subsection{Imaging}

NGC 6553 ($\alpha_{\rm J2000}$ = 18$^{\rm h}$09$^{\rm m}$17.6$^{\rm s}$,
$\delta_{\rm J2000}$ = $-$25$^{\rm o}$54$'$31$''$, l = 5.25$^{\circ}$, b =
$-$3.03$^{\circ}$) is located at low galactic latitude and is projected in the
direction of the Galactic centre with X = +5.9kpc, Y = +0.5kpc, and Z = $-$
0.3kpc. It has a relatively low extinction of E$_{B-V}$= 0.70 mag
(Guarnieri et al. 1998) and a high
luminosity of M$_{V}$ = $-$7.77 mag, (Harris 1996)\footnote[1]{\tt
http://www.phy\-sics.mc\-mas\-ter\-.ca/Glo\-bu\-lar.html}. 
 
The  $BVI$ observations of NGC 6553 were  obtained in June 2002 with the
wide-field imager (WFI) at the 2.2m ESO-MPI telescope (La Silla,
Chile). The data were obtained within our program dedicated to
surveying the Galactic globular clusters with the WFI (Zoccali et al. 2001). 
The $J$$H$$K_S$ colours are from the 2MASS Atlas
(Cutri et al. 2003)\footnote[2]{\tt
http://ipac.caltech.edu./2mass/releases/allsky/}. No J,H,K
photometry for the sample stars II-64 and 267092 are currently available.

\begin{figure}[ht]
\psfig{file=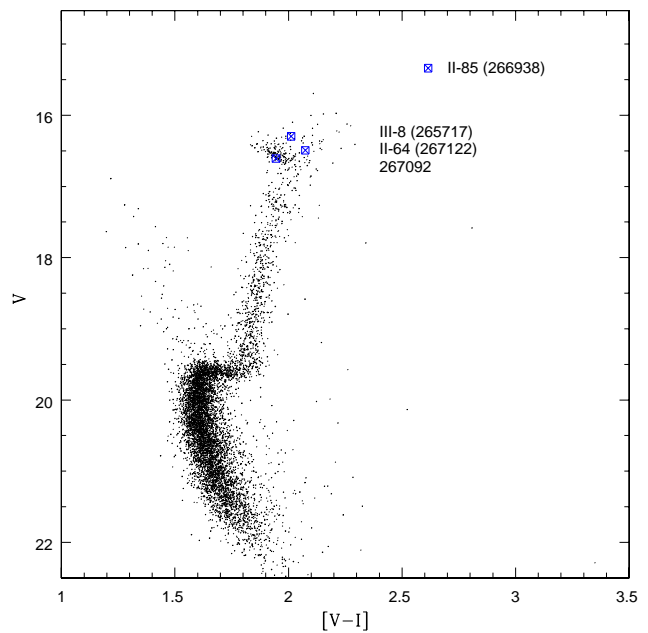,angle=0,width=9cm}
\caption {$V$ vs. $V-I$ CMD
 of NGC 6553 with the target stars indicated.}
\label{cmd}
\end{figure}

In Fig.~\ref{cmd} we show the location of target  sample stars  on the
colour-diagram magnitude (CMD) of NGC 6553 using Hubble Space Telescope (HST)
data (Ortolani et al. 1995), with three stars 
located at the horizontal branch (HB) level and one of
them at the tip of the RGB. The identifications follow the
notation given in Hartwick (1975)
 and in the WFI data obtained with the ESO-MPI
telescope. 

\subsection{High-resolution spectra}

 High-resolution spectra of four giants in NGC 6553, in the wavelength range
$\lambda\lambda$  4800-6800 {\rm \AA}, were  obtained with the UVES
spectrograph (Dekker et al. 2000) at the ESO VLT-UT2 Kueyen telescope. 
The red portion of the spectrum (5800-6800 {\rm \AA}) was obtained with the
MIT backside-illuminated ESO CCD \# 20, of
4096x2048 pixels, and pixel size  15x15$\mu$m.
The blue portion of the spectrum (4800-5800  {\rm \AA}) used ESO Marlene
EEV CCD-44, backside illuminated, 
4102x2048 pixels, and pixel size  15x15$\mu$m. 
With the UVES standard setup 580, the resolving power
(R=$\lambda/\Delta\lambda$) is 
 R $\simeq$ 55 000 for a slit of 0.8''. 
Typical signal-to-noise (S/N) ratios for the spectra were obtained 
considering average values at
different wavelengths. 
The pixel scale is 0.0147 {\rm \AA}/pix. 
Spectra of rapidly rotating hot B stars
 at similar airmasses as the target were also observed,
 in order to correct for telluric lines.  
Table \ref{logbook} shows the log of observations.

\begin{table*}
\caption[1]{Log of spectroscopic observations. {The S/N ratio is} given per resolution element (7 pixels);
the value per pixel  is the value per resolution element 
divided by $\sim$ 2.8.}
\label{logbook}
\begin{flushleft}
\centering
\begin{tabular}{ccccccccccccccc}
\noalign{\smallskip}
\hline \hline
\noalign{\smallskip}
\noalign{\vskip 0.1cm}
{\rm star} & $\alpha_{\rm J2000}$ & $\delta_{\rm J2000}$  &  date & UT & exp & Seeing  & Airmass & (S/N)/res.el.   \\  
           &  [h m s]               &  [d m s]               &        &    & [s] & $[\arcsec]$  &         &   \\
\noalign{\vskip 0.2cm}
\noalign{\hrule\vskip 0.2cm}
\noalign{\vskip 0.2cm}

II-64 267122 & 18:09:18.31 & -25:55:01.15 & 2000 June 26 & 14:21:25.00  & 2 $\times$ 3600 & 0.8 & 1.4& 110   \cr   
II-85 266938 & 18:09:17.96 & -25:55:04.89 & 2000 June 26 & 14:21:25.00  & 2 $\times$ 3600 & 0.8 & `` & 200   \cr    
III-8 265717 & 18:09:13.05 & -25:55:30.00 & 2000 June 26 & 17:37:49.00  & 2 $\times$ 5400 & 0.8 & 1.02 & 170   \cr
------267092 & 18:09:13.42 & -25:55:01.85 & 2000 June 26 & 17:37:49.00  & 2 $\times$ 5400  & 0.8 & 1.02& 110   \cr
\noalign{\smallskip} \hline \end{tabular}
\end{flushleft} 
\end{table*}

\begin{figure}[ht]
\psfig{file=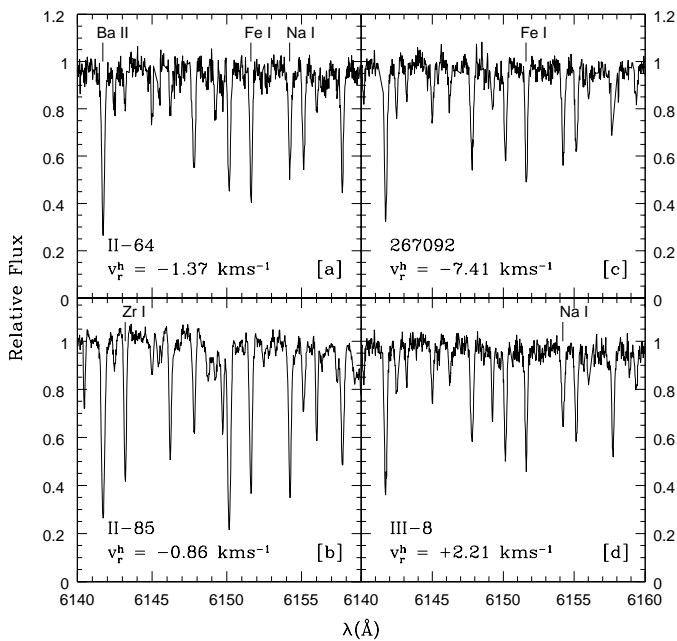,angle=0.,width=9.0 cm}
\caption{Typical raw spectra of NGC 6553 with some atomic
lines identified. In each panel the heliocentric radial velocity is given.}
\label{espec}
\end{figure}

The spectra were reduced using the UVES context of the MIDAS reduction package,
including bias and inter-order background subtraction, flatfield correction,
extraction and wavelength calibration (Ballester et al. 2000). 
The stars were observed in pairs, II-64 together with II-85 and
III-8 and 267092, and reduced taking this into account.
In Fig.~\ref{espec} typical sample spectra are shown. 

\subsection{Radial velocities}

The radial velocities v$_{\rm r}$ were measured with the automatic code {\tt
DAOSPEC} (Stetson \& Pancino 2006, in preparation) based on line shifts of a
given list of wavelengths in the range  5800$<$$\lambda$$<$6800 $\rm \AA$. 
The heliocentric radial velocities v$_{\rm r}^{\rm
h}$ were determined using the {\tt IRAF} task {\tt rvcorrect}.
With these methods, the standard errors of the
average velocities are $\sim$ 0.5 km s$^{-1}$. A mean v$_{\rm r}$ 
= $-$ 0.27 $\pm$ 1.99 ($\sigma$ = 3.98, 4 stars) km s$^{-1}$ or heliocentric v$^{\rm
hel}_{\rm r}$  = $-$ 1.86 $\pm$ 2.01 ($\sigma$ = 4.02, 4 stars) km s$^{-1}$ was found
for NGC 6553. Combining high-resolution radial velocities  of stars from
the present work and from Origlia et al. (2002), Barbuy et al. (1999), Cohen et al. (1999),
and Mel\'endez et al. (2003), a
mean value of v$_{\rm r}^{\rm h}$ = +1.30 $\pm$ 1.41 ($\sigma$ = 6.50 km
s$^{-1}$) was found. This result is in good agreement with the value of $-$1
km s$^{-1}$, based on different methods and from low-resolution spectra of Coelho et
al. (2001), and  the value 8.4 ($\sigma$ = +8.4, 21 stars) km s$^{-1}$ derived by
Rutledge et al. (1997).


\section{Stellar Parameters}

\subsection{Photometric parameters}

Adopting a reddening E$_{B-V}$ = 0.70 for NGC 6553
(Guarnieri et al. 1998; Sagar et al. 1999; Mel\'endez et al. 2003)
and reddening ratios  E$_{V-I}$/E$_{B-V}$=1.33
(Dean et al. 1978), E$_{V-K}$/E$_{B-V}$=2.744, and E$_{J-K}$/E$_{B-V}$=0.52
(Rieke \& Lebofsky 1985), we obtained dereddened colours to derive
photometric temperatures.
The 2MASS K$_s$ magnitudes were transformed to the TCS 
(Telescopio Carlos S\'anchez) system using relations by
Carpenter (2001) and Alonso et al. (1998). Effective photometric
temperatures were then determined based on the infrared flux method (IRFM)
 using the empirical transformations of Alonso et al. (1999, 2001, 
hereafter AAM99), where metallicity values provided by Mel\'endez et al. (2003)
were initially adopted. If we take into account the
1~$\sigma$ uncertainties provided by AAM99 on $V-I$ ($\sigma_{V-I}$ = 125
K), $V-K$ ($\sigma_{V-K}$ = 40 K), and $J-K$ ($\sigma_{J-K}$ = 125 K), the
 uncertainty in the photometric temperatures is about
130 K. Table~\ref{photo} lists the photometric data for the programme stars,
together with the temperature estimated from each colour. The final
photometric temperature for each star was averaged from these individual
values.

\begin{table}
\begin{flushleft}
\caption{Photometric effective temperatures T$_{\rm eff}$ derived using $V-I$,
$V-K$, and $J-K$ based on relations from AAM99. ($V-I$)$^{\rm C}_{0}$ and
($V-I$)$^{\rm J}_{0}$ refer to reddening-corrected
 Johnson-Cousins and Johnson photometric systems.}             
\label{photo}      
\centering          
\begin{tabular}{lccccc}     
\noalign{\smallskip}
\hline\hline    
\noalign{\smallskip}
\noalign{\vskip 0.1cm} 
Parameter & II-64 & II-85 & III-8 & 267092 \\                 
\noalign{\vskip 0.1cm}
\noalign{\hrule\vskip 0.1cm}
\noalign{\vskip 0.1cm}    
\multicolumn{5}{c}{\hbox{Magnitudes and Colours}} \\
\noalign{\vskip 0.1cm}
\noalign{\hrule\vskip 0.1cm}
\noalign{\vskip 0.1cm} 
V [mag]       & 16.492 & 15.339 & 16.297 & 16.608 \\ 
I [mag]       & 14.418 & 12.724 & 14.285 & 14.663 \\ 
J [mag]       &  ---   & 10.979 & 12.798 &  ---   \\ 
H [mag]       &  ---   &  9.975 & 11.983 &  ---   \\ 
K$_{S}$ [mag] &  ---   &  9.647 & 11.749 &  ---   \\ 
($V-I$)$^{\rm C}_{0}$  &  1.143   &  1.684 & 1.081 &  1.014   \\
($V-I$)$^{\rm J}_{0}$  &  1.469   &  2.172 & 1.389 &  1.303   \\
\noalign{\vskip 0.1cm}
\noalign{\hrule\vskip 0.1cm}
\noalign{\vskip 0.1cm}    
\multicolumn{5}{c}{\hbox{Effective Temperatures}} \\
\noalign{\vskip 0.1cm}
\noalign{\hrule\vskip 0.1cm}
\noalign{\vskip 0.1cm}  
\hbox{T$_{\rm V-I}$ [K]} & 4448 & 3773 & 4564 & 4700 \\
\hbox{T$_{\rm V-K}$ [K]} &  --- & 3853 & 4500 &  --- \\
\hbox{T$_{\rm J-K}$ [K]} &  --- & 3901 & 4564 &  --- \\
\hbox{$T_{mean}$ [K]}    & 4448 & 3842 & 4543 & 4700 \\
\hline                  
\end{tabular}
\end{flushleft}
\end{table}  

The surface gravities $\log g$ were
derived with the classical relation

\begin{displaymath}
\log({\frac{g_{*}}{ g_{\odot}}}) = + 4\log (\frac{T_*}{T_{\odot}}) +
0.4(M_{\rm bol}^{*} - M_{\rm bol}^{\odot}) +  \log (\frac{M_*}{M_{\odot}}) 
\end{displaymath}

\noindent by adopting  $T_{\odot}$ = 5780 K, M$_{\rm bol^{\odot}}$ =
4.75, $\log g_{\odot}$ = 4.44 dex, and M$_*$ = 0.80 M$_{\odot}$.
To determine $M_{\rm bol}^{*}$ we 
assume a distance modulus of $(m-M)_{\rm 0}$ = 13.60,
visual selective-to-total extinction 
A$_V$ = 2.17, and bolometric correction $BC_V$ values from 
AAM99.

We adopt errors due to uncertainties in the photometric
effective temperature of T$_{\rm
eff}$ ($\sigma_{T_{\rm eff}}$ = 130 K, according to AAM99)
and in the stellar mass of
M$_{*}$ ($\sigma_{M_{*}}$= 0.2M$_{\odot}$).
 The error in 
 M$_{bol}$  = (M$_{\rm V}$ - A$_{\rm V}$) + BC(V) is mostly
due to the M$_{\rm V}$ value and
 total extinction A$_{\rm V}$, the latter with an uncertainty
around $\pm$0.05 mag. For M$_{\rm V}$, adopting an error in
distance as large as 20\%, we get  $\sigma_{M_{\rm bol}}$ $\sim$ 0.2  mag,
and  errors on the adopted photometric gravities  of $\pm$0.2
dex.

\subsection{Spectroscopic parameters}

\subsubsection{Equivalent widths, oscillator strengths, and damping
constants}

Equivalent widths W$_{\lambda}$ of \ion{Fe}{I} and
\ion{Fe}{II} lines were measured using the automatic code {\tt
DAOSPEC} developed by P. Stetson \& E. Pancino (2006, in
preparation). Given a reference line list 
and a first estimate of FWHM by an iterative process, {\tt DAOSPEC} fits a
continuum and measures  W$_{\lambda}$ by fitting a Gaussian profile of
fixed width. The overall uncertainty is $<$ 10\%. A comparison of
W$_{\lambda}$s measured with both {\tt IRAF} and 
{\tt DAOSPEC} was presented in
Fig.~3 of Alves-Brito et al. (2005). Lines with
 10$<$ W$_{\lambda}$ $<$ 150 m$\rm \AA$ were selected.

The oscillator strengths $\log gf$ of FeI lines were taken from
the National Institute of Standards \& Technology (NIST) atomic
database (Martin et al. 2002), while those of the Fe II lines 
were adopted according to the renormalized values 
from Mel\'endez \& Barbuy (2002; 2006, in preparation) and Mel\'endez et al.
(2006).
The Ba II 6141, 6496, La II 6390, and Eu II 6645 ${\rm \AA}$ lines
show a hyperfine splitting structure, and those were taken into
account when adopting data from Biehl (1976) and Lawler (2001),
together with solar isotopic ratios.   
Damping constants and oscillator strengths for the elements
other than Fe were adopted from Barbuy et al. (2006).
Solar abundances were adopted from Grevesse \& Sauval (1998).

\subsubsection{Derivation of spectroscopic parameters}

The effective temperatures were then checked by imposing an excitation
equilibrium
on FeI and FeII lines of different excitation potentials.
The FeI and FeII lines employed in this analysis are listed in Table A.2,
together with their equivalent widths for the sample stars.
 Hereafter we refer to the atmospheric parameters [$T_{\rm eff}$, $v_{t}$, $\log g$] obtained by
imposing constraints on Fe I and/or Fe II abundances as 
spectroscopic parameters. 
A change of 100 K on $T_{\rm eff}$  causes a
recognisable trend in the plane FeI abundance versus excitation potential, 
such that this can be considered as a reasonable  uncertainty 
value. The microturbulent velocities $v_{t}$  were determined using FeI lines.
 The uncertainty derived from the  FeI abundance versus W$_{\lambda}$
is  0.2 km s$^{-1}$.

\begin{figure}[ht]
\psfig{file=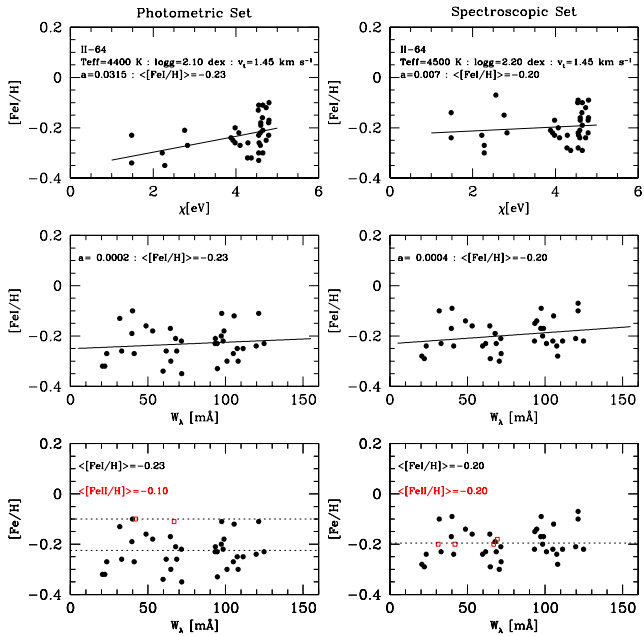,angle=0,width=9cm}
\caption {Demonstration of the offset in temperature, surface gravity, and
metallicity found for the star II-64 based on photometric and spectroscopic
methods. The constant $\tt a$ is the slope of the least-squares fit
to the data. $\it Top$: FeI vs. $\chi$. $\it Middle$: FeI vs. W$_{\lambda}$. $\it
Bottom$: FeI ($\it circles$) and FeII ($\it squares$) vs. W$_{\lambda}$. The
dotted line corresponds to the mean values found using only those measured
within 1~$\sigma$, as employed by the $\sigma$ clipping method.}
\label{ii64}
\end{figure}

The spectroscopic gravity $\log g$ was derived from ionization equilibrium of FeI and FeII lines.
 We found a mean offset between the spectroscopic and photometric (Sect. 3.2.1)
values of $\Delta \log g$ (spectro - photo) = +0.17 $\pm$ 0.08
 ($\sigma$ = +0.17, 4 stars). However, this procedure
is affected by several uncertainties,  amounting to an upper limit  
of $\pm$ 0.30 dex. This value corresponds to a difference higher than 1~$\sigma$ 
 in the value of [FeII/H]. In Fig.~\ref{ii64}  an example of
 determination of the atmospheric parameters for the star II-64 is shown. 

\subsection{Final parameters and metallicity}

To derive metallicities,  Fe I and Fe II lines
with 15 $<$ W$_{\lambda}$ $<$ 150 m$\rm\AA$ and $\lambda >$ 5800
$\rm\AA$ were selected. Photospheric 1D models for the sample giants 
were extracted from the {\tt NMARCS} grid (Plez et al. 1992).

The LTE abundance analysis and the spectrum synthesis calculations were
performed using the codes by Spite (1967) 
and subsequent improvements in the last
thirty years, 
described in Cayrel et al. (1991) and Barbuy et al. (2003).

Stellar parameters were derived by initially adopting the photometric
effective temperature and gravity and  then further constraining the
temperature by imposing excitation equilibrium for \ion{Fe}{I} lines
and the gravity by imposing ionization equilibrium for \ion{Fe}{I} and
\ion{Fe}{II}, as described in previous sections. The entire process of defining
the atmospheric parameters is iterated until a consistent set of model
atmosphere parameters is finally obtained. Table ~\ref{final} summarises
the derived atmospheric parameters. We adopt the
spectroscopic parameters for the subsequent analysis, 
having used the photometric parameters
only as initial values; therefore, the  uncertainties on the reddening 
described in Sect. 3.1 do not affect the final results.
We point out otherwise that the good agreement between the spectroscopic
and photometric parameters indicates that the reddening value
adopted is satisfactory.

\begin{table*}
\begin{flushleft}
\caption{Photometric and adopted spectroscopic stellar parameters}             
\label{final}      
\centering          
\begin{tabular}{ccccccc}     
\noalign{\smallskip}
\hline\hline    
\noalign{\smallskip}
\noalign{\vskip 0.1cm} 
Star & M$_{\rm bol}$  & $T_{\rm eff}$ & $v_{t}$ & $\log g^{\rm a}$ & [FeI/H] & [FeII/H] \\                 
     & mag & K & km s$^{-1}$ & dex & dex & dex \\
\noalign{\vskip 0.1cm}
\noalign{\hrule\vskip 0.1cm}
\noalign{\vskip 0.1cm}    
\multicolumn{7}{c}{\hbox{Photometric}} \\
\noalign{\vskip 0.1cm}
\noalign{\hrule\vskip 0.1cm}
\noalign{\vskip 0.1cm}
II-64 267122 & +0.1987   & 4448  & 1.45   & 2.07  & $-0.23$ $\pm$ 0.15 & $-0.10$ $\pm$ 0.03  \\ 
II-85 266938 & $-1.6190$ & 3842  & 1.38   & 1.08  & $-0.23$ $\pm$ 0.15 & $-0.29$ $\pm$ 0.09 \\
III-8 265717 & +0.0653   & 4543  & 1.32   & 2.05  & $-0.21$ $\pm$ 0.16  & $-0.23$ $\pm$ 0.09  \\
267092 & +0.4672   & 4700          & 1.50   & 2.27  & $-0.19$ $\pm$ 0.11 & $-0.46$ $\pm$ 0.06  \\
\noalign{\vskip 0.1cm}
\noalign{\hrule\vskip 0.1cm}
\multicolumn{7}{c}{\hbox{Spectroscopic}}  \\ 
\noalign{\vskip 0.1cm}
\noalign{\hrule\vskip 0.1cm}    
II-64 267122 & --- & 4500 & 1.45 & 2.20 & $-0.20$ $\pm$ 0.15  & $-0.20$ $\pm$ 0.01  \\ 
II-85 266938 & --- & 3800 & 1.38 & 1.10 & $-0.23$ $\pm$ 0.15  & $-0.29$ $\pm$ 0.09  \\
III-8 265717 & --- & 4600 & 1.40 & 2.40 & $-0.17$ $\pm$ 0.15  & $-0.17$ $\pm$ 0.04   \\
267092       & --- & 4600 & 1.50 & 2.50 & $-0.21$ $\pm$ 0.12  & $-0.22$ $\pm$ 0.06  \\
\hline                  
\end{tabular}
\end{flushleft}
\end{table*}

The mean offset in the $T_{\rm eff}$ values is 25 $\pm$ 48 K ($\sigma$ = 96, 4 stars)
in the sense of spectroscopic minus photometric temperature.
This means that the mean  $T_{\rm eff}$   based on photometry is  $\sim$25~K cooler 
than the spectroscopic ones.  
Along the iterative process of [Fe/H] determination,
 discrepant lines were clipped out within 1~$\sigma$ of the mean.
Tests on the sample stars showed that
a 2~$\sigma$ clipping gives differences of [Fe/H]$<\pm$0.01.
 A mean metallicity of [Fe/H] = $-0.20$ $\pm$ 0.01 ($\sigma$ = 0.02, 4 stars)
is found for the sample stars. 
The spectroscopic parameters were adopted for the abundance analysis.


\section{Spectrum synthesis} 

Abundance ratios were derived for the following eleven elements Na,
Mg, Al, Si, Ca, Ti, Fe, Zr, Ba, La, and Eu.
Most lines used are the same reported in Barbuy et al. (2006) as those
Table~A.1 gives the relevant parameters used in
the spectrum synthesis of each line and resulting abundances.

Elemental abundances were obtained through line-by-line spectrum synthesis
calculations with atomic lines as described in Sect. 3
and molecular lines of CN
A$^2$$\Pi$-X$^2$$\Sigma$, C$_2$  Swan  A$^3$$\Pi$-X$^3$$\Pi$ and TiO
A$^3$$\Phi$-X$^3$$\Delta$ $\gamma$ and B$^3$$\Pi$-X$^3$$\Delta$ $\gamma$'
systems  taken into account. 
Table~\ref{meanfinal} gives the final mean results for each star, where the
abundances were normalized to FeI for neutral elements and FeII for ionized ones.

\begin{figure}[ht]
\psfig{file=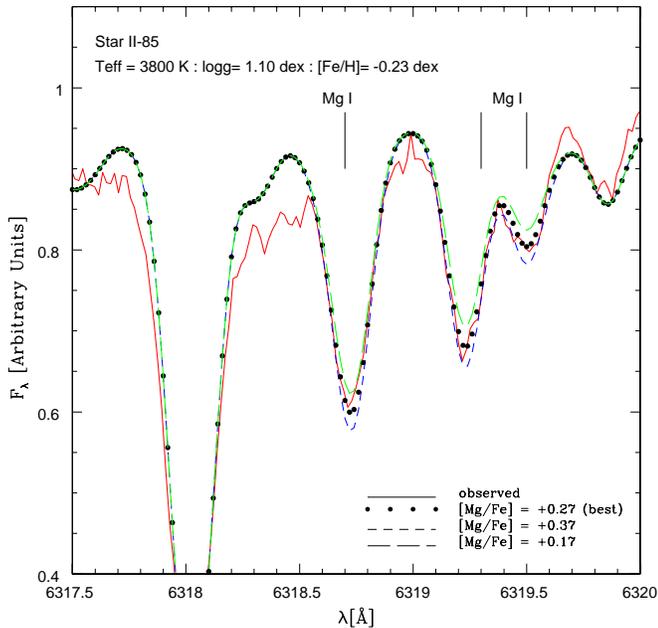,angle=0.,width=9.0 cm}
\caption{Magnesium triplet at 6318.7, 6319.2, 6319.4  ${\rm \AA}$ in
II-85 computed with [Mg/Fe] = +0.27 ({\it dotted line}, best fit), 
[Mg/Fe] = +0.37 ({\it short-dashed line}), and [Mg/Fe] = +0.17 ({\it long-dashed line}). }
\label{mg}
\end{figure}

\begin{figure}[ht]
\psfig{file=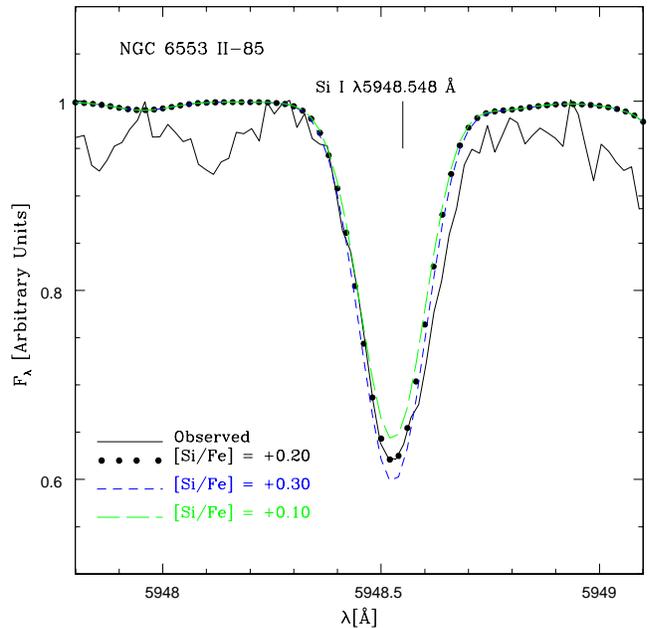,angle=0.,width=9.0 cm}
\caption{Silicon line at 5948.548  ${\rm \AA}$ in
II-85 computed with [Si/Fe] = +0.20 ({\it dotted line}, best fit), 
[Si/Fe] = +0.30 ({\it short-dashed line}), and [Si/Fe] = +0.10 
({\it long-dashed line}).}
\label{siII85novo}
\end{figure}

\begin{figure}[ht]
\psfig{file=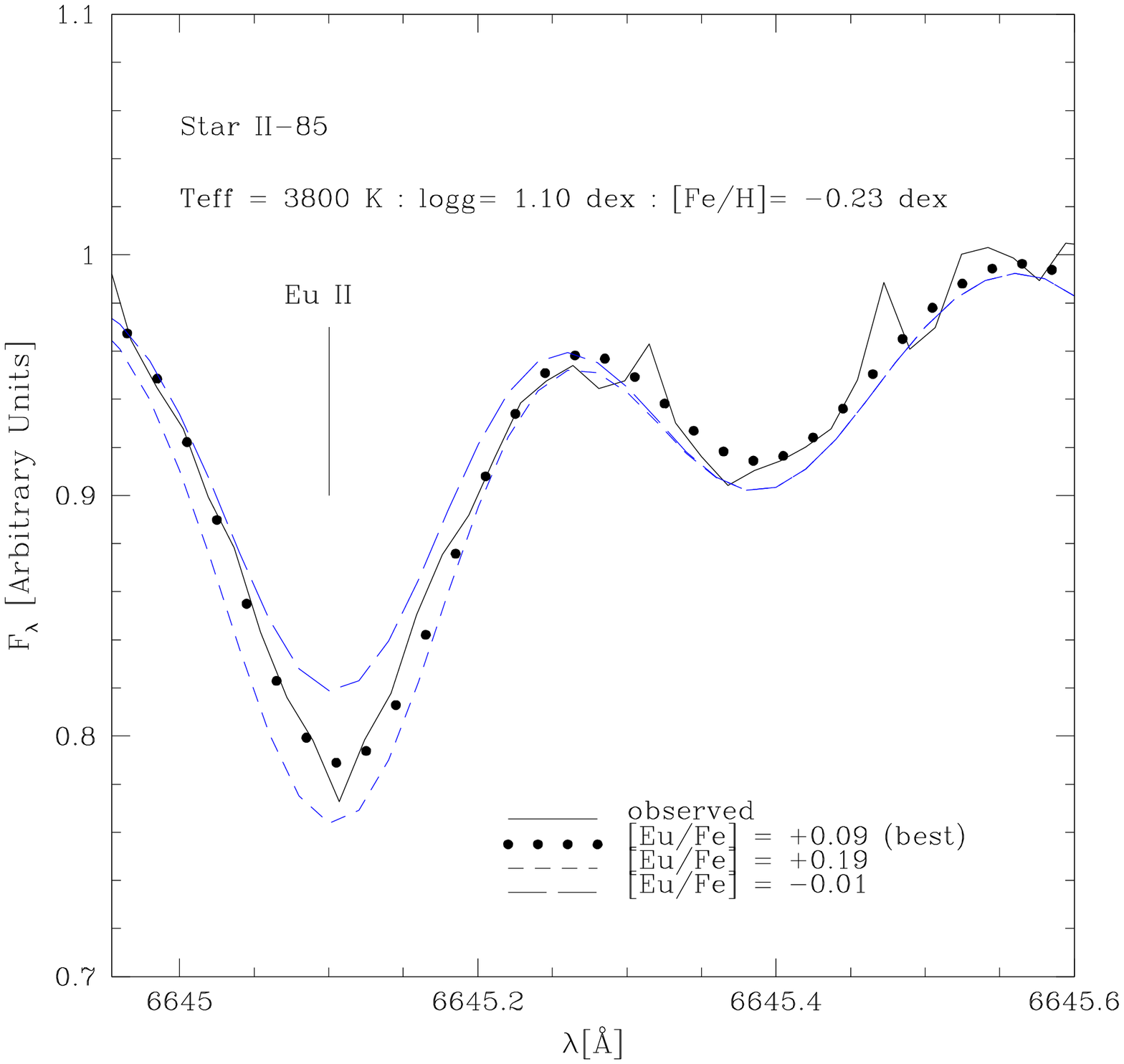,angle=0.,width=9.0 cm}
\caption{Europium line at 6645 ${\rm \AA}$ in
II-85 computed with [Eu/Fe] = +0.09 ({\it dotted line}, best fit), 
[Eu/Fe] = +0.19 ({\it short-dashed line}), and [Eu/Fe] = $-$0.01 
({\it long-dashed line}). }
\label{eu}
\end{figure}

Abundances of the $\alpha$-elements Mg,
Ca, Si, and Ti were derived. 
The oxygen forbidden line at 6300 ${\rm \AA}$ 
is blended with the strong sky emission line,
given the low radial velocity of the cluster (Sect. 2.3),
cannot be studied. In a previous work, Mel\'endez et al. (2003)
derived the oxygen abundance of 5 giants in NGC 6553, based
on OH infrared lines. We adopted the carbon, nitrogen, and
oxygen abundances from that work in the present calculations,
namely [C/Fe]=$-$0.7, [N/Fe]= +1.30, [O/Fe]=+0.25.
Figures \ref{mg} and \ref{siII85novo} show the magnesium triplet at
 6318.7, 6319.2, 6319.4 ${\rm \AA}$, and SiI 5948.548 ${\rm \AA}$ in
the giant II-85.
Figure \ref{eu} shows the Europium line at 6645.127  ${\rm \AA}$ in
II-85. 

\begin{table*}
\begin{flushleft}
\caption{Mean abundance ratios of the programme stars and final mean cluster abundances derived}             
\label{meanfinal}      
\centering          
\begin{tabular}{ccccccccccccccc}     
\noalign{\smallskip}
\hline\hline    
\noalign{\smallskip}
\noalign{\vskip 0.1cm} 
\hbox{}	& \hbox{} & \multicolumn{2}{c}{II-64} & & \multicolumn{2}{c}{II-85} && \multicolumn{2}{c}{III-8} && \multicolumn{2}{c}{267092} & mean   \\
\cline{3-4} \cline{6-7} \cline{9-10} \cline{12-13}  \\
\hbox{Species} & \hbox{$\epsilon_{\odot}$(X)$^{\mathrm{a}}$} & \hbox{N} & \hbox{[X/Fe]} && \hbox{N} & \hbox{[X/Fe]} && \hbox{N} & \hbox{[X/Fe]} && \hbox{N} & \hbox{[X/Fe]} & \hbox{[X/Fe]} & \\
\noalign{\smallskip}
\noalign{\smallskip}
\noalign{\vskip 0.1cm}
\noalign{\hrule\vskip 0.1cm}
\multicolumn{14}{c}{\hbox{$\alpha$-elements}}  \\ 
\noalign{\vskip 0.1cm}
\noalign{\hrule\vskip 0.1cm}   
\hbox{Mg I} & \hbox{7.58} & \hbox{3} & \hbox{+0.32 $\pm$ 0.09} && \hbox{3} & \hbox{+0.27} && \hbox{3} & \hbox{+0.29 $\pm$ 0.03} && \hbox{3} &
\hbox{+0.22 $\pm$ 0.09} & \hbox{+0.28 $\pm$ 0.04}  \\
\hbox{Si I} & \hbox{7.55} & \hbox{6} & \hbox{+0.27 $\pm$ 0.09} && \hbox{4} & \hbox{+0.22 $\pm$ 0.08} && \hbox{6} & \hbox{+0.17 $\pm$ 0.07} &&
\hbox{6} & \hbox{+0.17 $\pm$ 0.03}  & \hbox{+0.21 $\pm$ 0.05}  \\
\hbox{Ca I} & \hbox{6.36} & \hbox{15} & \hbox{$+$0.11 $\pm$ 0.12} && \hbox{14} & \hbox{$+$0.12 $\pm$ 0.09} && \hbox{15} & \hbox{$+$0.05 $\pm$ 0.14}
 && \hbox{15} &
\hbox{$-$0.07 $\pm$ 0.13}  & \hbox{+0.05 $\pm$ 0.09}   \\
\hbox{Ti I} & \hbox{5.02} & \hbox{15} & \hbox{$+$0.04 $\pm$ 0.08} && \hbox{15} &
\hbox{+0.15 $\pm$ 0.11} && \hbox{15} & \hbox{$-$0.05 $\pm$ 0.10} && \hbox{15} & \hbox{$-$0.17 $\pm$ 0.15}   & \hbox{$-$0.01 $\pm$ 0.14}  \\
\hbox{Ti II} & \hbox{5.02} & \hbox{2} & \hbox{$+$0.06 $\pm$ 0.04} && \hbox{2} & \hbox{$+$0.01 $\pm$ 0.11} && \hbox{2} & \hbox{$-$0.04 $\pm$
0.04} && \hbox{2} & \hbox{$-$0.07}  & \hbox{$-$0.02 $\pm$ 0.06}  \\

\noalign{\vskip 0.1cm}
\noalign{\hrule\vskip 0.1cm}
\multicolumn{13}{c}{\hbox{Odd-Z elements}}  \\ 
\noalign{\vskip 0.1cm}
\noalign{\hrule\vskip 0.1cm}   
\hbox{Na I} & \hbox{6.33} & \hbox{2} & \hbox{+0.12} && \hbox{3} & \hbox{+0.34} && \hbox{2} & \hbox{$-$0.08} && \hbox{2} & \hbox{+0.12}
 & \hbox{+0.16 $\pm$ 0.23}  \\
\hbox{Al I} & \hbox{6.47} & \hbox{2} & \hbox{+0.25 $\pm$ 0.03} && \hbox{---} & \hbox{---} && \hbox{2} & \hbox{+0.15 $\pm$ 0.17} && \hbox{2} &
\hbox{+0.13}  & \hbox{+0.18 $\pm$ 0.06}  \\

\noalign{\vskip 0.1cm}
\noalign{\hrule\vskip 0.1cm}
\multicolumn{13}{c}{\hbox{Heavy elements}}  \\ 
\noalign{\vskip 0.1cm}
\noalign{\hrule\vskip 0.1cm}   
\hbox{Zr I} & \hbox{2.60} & \hbox{1} & \hbox{$-0.74$} && \hbox{1} & \hbox{$- 0.64$} && \hbox{1} & \hbox{$- 0.64$} && \hbox{1} & \hbox{$- 0.64$} 
 & \hbox{$-$0.67 $\pm$ 0.05} \\
\hbox{Ba II} & \hbox{2.13} & \hbox{2} & \hbox{$-$0.08} && \hbox{2} & \hbox{$-$0.18 $\pm$ 0.14} && \hbox{2} & \hbox{$-$0.48 $\pm$ 0.14} && \hbox{2} & 
\hbox{$-$0.38}  & \hbox{$-$0.28 $\pm$ 0.18}   \\
\hbox{La II} & \hbox{1.17} & \hbox{1} & \hbox{+0.01} && \hbox{1}
 & \hbox{$-0.22$} && \hbox{---} & \hbox{---} && \hbox{---} & \hbox{---}   & \hbox{$-$0.11 $\pm$ 0.16}  \\
\hbox{Eu II} & \hbox{0.51} & \hbox{1} & \hbox{+0.09} && \hbox{1} & \hbox{+0.09} && 
\hbox{1} & \hbox{+0.10} && \hbox{1} & \hbox{+0.14}  & \hbox{+0.10 $\pm$ 0.02}  \\ 
\noalign{\vskip 0.1cm}
\noalign{\hrule\vskip 0.1cm}
\multicolumn{13}{c}{\hbox{[Fe/H]}}  \\ 
\noalign{\vskip 0.1cm}
\noalign{\hrule\vskip 0.1cm}   
\hbox{Fe I} & \hbox{7.50} & \hbox{56} & \hbox{$-0.20$ $\pm$ 0.15} && \hbox{52}
& \hbox{$-0.23$ $\pm$ 0.15} && \hbox{59} & \hbox{$-0.17$ $\pm$ 0.15} && \hbox{57} & \hbox{$-0.21$ $\pm$ 0.12}  & \hbox{$-$0.20 $\pm$ 0.02}  \\
\hbox{Fe II} & \hbox{7.50} & \hbox{4} & \hbox{$-0.20$ $\pm$ 0.01} && \hbox{2} &\hbox{$-0.29$ $\pm$ 0.09} && 
\hbox{3} & \hbox{$-0.17$ $\pm$ 0.04}
&& \hbox{4} &\hbox{$-0.22$ $\pm$ 0.06}  & \hbox{$-0.22$ $\pm$ 0.05}  \\
\hline                 
 \end{tabular}
\end{flushleft}
\begin{list}{}{}
\item[$^{\mathrm{a}}$] Solar abundances are from Grevesse \& Sauval 
(1998).
\end{list}
\end{table*}

\subsection{Errors in abundance ratios}

Abundances for each species given in Table 4 were obtained
by averaging the abundances resulting from all measured lines. Consequently, we
give the 1~$\sigma$ scatter (standard deviation) of each measurement, such that the
standard error can be obtained easily.

The uncertainties on atmospheric parameters were discussed in
Sect. 3.
To estimate the uncertainties on the derivation of abundances 
due to the choice of stellar parameters, we show the sensitivity of the
abundances in Table \ref{error} for the star II-64
by varying the temperature by 100~K, surface gravity by +0.30 dex, and 
microturbulent velocity by +0.20 km s$^{-1}$, which are typical errors
in our analysis. The total error is given in the last column as
the quadratic sum of all uncertainties. We can see that the total uncertainties
 are lower than 0.21 dex. In the next sections these
uncertainties correspond to the 1~$\sigma$ scatter for each atomic species.

\begin{table*}
\begin{flushleft}
\centering
\caption{Sensitivity of abundances to changes in
$\Delta$T$_{\rm eff}$ = 100 K,
$\Delta$log g = +0.3, and $\Delta$v$_{\rm t}$ = 0.2 km s$^{-1}$. In the last
column the corresponding total error is given. } 
\label{error}

\begin{tabular}{lccccc}
\hline\hline
\noalign{\smallskip}
\hbox{Species}  & \hbox{$\Delta$T} & \hbox{$\Delta$$\log$ g} & \hbox{$\Delta$v$_{t}$} & \hbox{($\sum$x$^{2}$)$^{1/2}$} \\
\hbox{(1)} & \hbox{(2)}& \hbox{(3)} & \hbox{(4)} & \hbox{(5)} \\
\noalign{\smallskip}
\hline
\noalign{\smallskip}
\noalign{\vskip 0.1cm}
\noalign{\hrule\vskip 0.1cm}
\noalign{\vskip 0.1cm}
\multicolumn{5}{c}{\bf II-64}\\
\noalign{\vskip 0.1cm}
\noalign{\hrule\vskip 0.1cm}
\hbox{[NaI/Fe]}    & +0.08   & +0.02   & $-$0.02 & +0.08 \\
\hbox{[MgI/Fe]}    &  $-$0.01   &  $-$0.02   & +0.01   & +0.02 \\
\hbox{[AlI/Fe]}    & + 0.10   & +0.00   & +0.01   & +0.10 \\
\hbox{[SiI/Fe]}    &  $-$0.10   &  $-$0.01 & $-$0.05 & +0.11 \\
\hbox{[CaI/Fe]}    & +0.05   & +0.01   & $-$ 0.15 & +0.16 \\
\hbox{[TiI/Fe]}    & +0.10   & +0.03   & $-$ 0.10 & +0.14 \\
\hbox{[TiII/Fe]}   & +0.05   & +0.20   & +0.05   & +0.21 \\
\hbox{[ZrI/Fe]}    & +0.20   & +0.01   & +0.01   & +0.20 \\
\hbox{[BaII/Fe]}   & +0.01   & +0.05   & $-$ 0.20 & +0.21 \\
\hbox{[LaII/Fe]}   & +0.01   & +0.05   & +0.02   & +0.05 \\
\hbox{[EuII/Fe]}   & +0.01   & +0.06   & +0.01   & +0.06 \\
\hline
\noalign{\smallskip}
\noalign{\vskip 0.1cm}
\noalign{\hrule\vskip 0.1cm}
\noalign{\vskip 0.1cm}
\hbox{[FeI/H]}    & $-$0.03 & $-$0.04	& +0.07   & +0.09 \\
\hbox{[FeII/H]}   & +0.12 & $-$0.11  & +0.06   & +0.17 \\
\hline
\end{tabular}
\end{flushleft}
\end{table*}


\subsection{Comparison with previous work}

In the present work a metallicity of [Fe/H]=$-$0.20 is found for NGC 6553.
A derivation based on photometric parameters gives [FeI/H]=$-$0.22
and [FeII/H]=$-$0.21, whereas for star 267092 it gives  [FeI/H]=$-$0.19 
and [FeII/H]=$-$0.46.
In other words, the metallicities are the same whether they are computed
with the photometric or the spectroscopic parameters, since the two sets
of parameters are similar; and as demonstrated in Fig. 3, the spectroscopic
parameters give a better agreement between the four stars than
the photometric ones.

\subsubsection{Comparison with Barbuy et al. (1992)}

In Barbuy et al. (1992) a metallicity of [Fe/H]=$-$0.20 was found,
in agreement with the present work.
 A reddening 
E(B-V)=0.73 and a photometric effective temperature was adopted.
 Despite the much lower S/N of those spectra, the overall fit 
of synthetic spectra, which take blanketing 
lowering the continuum into account, proved to be a satisfactory method.

\subsubsection{Comparison with Barbuy et al. (1999)}

The atmospheric
parameters adopted for  II-85, the single star in common with
the present work,
were  T$_{\rm eff}$=4000 K, 
log g = 1.26 (0.70), v$_{\rm t}$=1.30 km s$^{-1}$ and
[Fe/H]=$-$0.50. The VI colours obtained with HST (Ortolani 
et al. 1995) were used, which differ slightly from the
WFI ones used in the present paper.

We found 3 main sources of uncertainty in the metallicity 
of [Fe/H]=$-$0.55 found in Barbuy et al. (1999),
 by comparing the parameters of star II-85 in
common with the present work. 

(i) The effective temperature of 4000 K was derived
in Barbuy et al. (1999) using photometry, adopting
a reddening of E(B-V)=0.70,
whereas in  the present work a
spectroscopic excitation temperature of 3842 K is obtained. 

(ii) The main source of discrepancy has been an
inappropriate correction of gravity by 0.60 dex.
The difference of log g = 1.26 found then from classical
formulae and log g = 0.70 as adopted led to a
$\Delta$[Fe/H]=$-$0.3.

\begin{figure}[ht]
\psfig{file=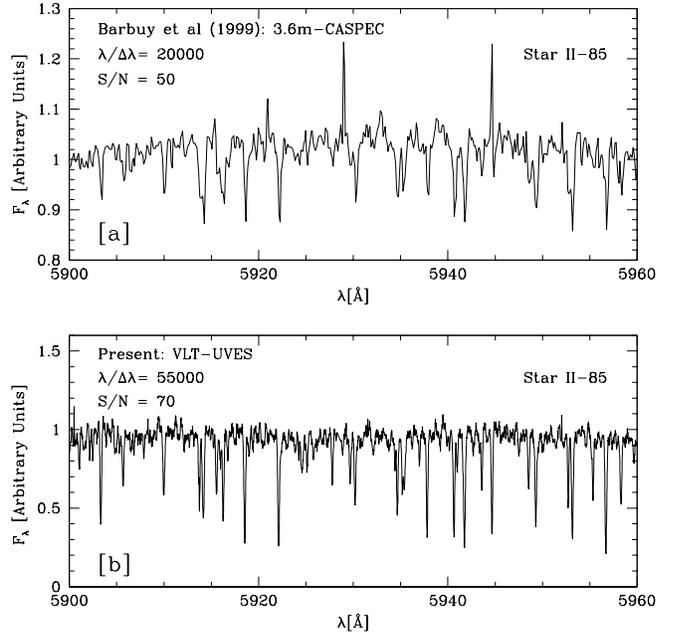,angle=0.,width=9.0cm}
\caption{Spectra of the giant II-85: comparison between the spectrum
studied in Barbuy et al. (1999),
obtained with the 3.6m telescope and the CASPEC spectrograph
(R $\sim$ 20 000)
and the present spectrum obtained with VLT-UVES (R$\sim$55 000).}
\label{especcomp}
\end{figure}

(iii) Equivalent widths for the star II-85 for lines in common are
compared in Table \ref{fex}. 
It appears that in the early work of 1999, W$_{\lambda}$'s of strong lines were
systematically lower by about 20 m${\rm \AA}$, 
which leads to a metallicity that is lower by $\Delta$[Fe/H] $\approx$ 0.1 dex. 
The reason for the discrepancy in  W$_{\lambda}$'s comes from the lower quality
of the 1999 CASPEC spectra both in resolution and S/N. 
This can be seen in Fig. \ref{especcomp}, which compares
 the two sets of observations for the star II-85.

\begin{table*}
\caption[1]{FeI lines used in the present work in common with Barbuy et al. (1999).}
\label{fex}
\begin{flushleft}
\centering
\begin{tabular}{ccccccccccccccc}
\noalign{\smallskip}
\hline \hline
\noalign{\smallskip}
\noalign{\vskip 0.1cm}
${\lambda ({\rm \AA})}$ & ${\chi_{\rm ex}}$(eV) & log gf  &  Present & Barbuy et al. 1999 & \\  
\noalign{\vskip 0.1cm}
\noalign{\hrule\vskip 0.1cm}
\noalign{\vskip 0.1cm}                  
5861.110 & 4.280 &    -2.450   &            24.9  &     25. \\
6082.710 & 2.220 &    -3.570   &           131.1  &    102. \\
6096.660 & 3.980 &    -1.930   &            70.1  &     74. \\
6120.250 & 0.910 &    -5.950   &            96.4  &     69. \\
6475.620 & 2.560 &    -2.940   &           127.5  &    118. \\
6481.870 & 2.280 &    -2.980   &           148.5  &    117. \\
6574.230 & 0.990 &    -5.040   &           149.3  &    130. \\
6739.520 & 1.560 &    -4.940   &            98.4  &    49. \\
\noalign{\smallskip} \hline \end{tabular}
\end{flushleft} 
\end{table*}


\subsubsection{Comparison with Cohen et al. (1999)}

The five stars of Cohen et al. were red horizontal branch
stars with T$_{\rm eff}$=4630-4830 K, 
log g=2.3, v$_{\rm t}$=1.4-2.5 km s$^{-1}$,
and a mean metallicity of [FeI/H]=-0.16
and  [FeII/H]=$-$0.18.
Effective temperatures were adopted from
excitation equilibrium of FeI lines. In order
to be compatible with the colours, they deduced
E(B-V)=0.78 for 3 stars and E(B-V)=0.86 for the
other two stars.
 The metallicity is in good agreement
with the present work, whereas there is some discrepancy
concerning the abundance ratios, as shown in Table \ref{fecomp}.
A difference $\Delta$[Ca/Fe]=0.2 is explained by an offset in the solar
Ca abundance adopted (Table \ref{fecomp}), whereas the oscillator strengths for the lines
in common agree. No lines in common are available for comparing
Ti oscillator strengths, but there is an offset in the solar abundance
of $\Delta$[Ti/Fe]=0.05.
Possible differences in damping constants cannot be checked since they
are not given in Cohen et al. (1999).

\begin{table*}
\caption[1]{Comparison of solar abundances and abundance ratios obtained
for NGC 6553 between the present work and Cohen et al. (1999).}
\label{fecomp}
\begin{flushleft}
\centering
\begin{tabular}{ccccccccccccccc}
\noalign{\smallskip}
\hline \hline
\noalign{\smallskip}
\noalign{\vskip 0.1cm}
 &   \multicolumn{2}{c}{\hbox{\bf Present work}}  & \multicolumn{2}{c}{\hbox{\bf Cohen et al. 1999}} \\ 
\noalign{\hrule\vskip 0.1cm}
Element  & $\epsilon$$_{\rm \odot}$(X) & [X/Fe] & $\epsilon$$_{\rm \odot}$(X)
 & [X/Fe] & \\  
\noalign{\vskip 0.1cm}
\noalign{\hrule\vskip 0.1cm}
\noalign{\vskip 0.1cm}                
O &  8.77  &  +0.2 &  8.68 & +0.50 \\ 
Mg & 7.58  &  +0.28 &  7.54 & +0.41 \\
Si & 7.55 &   +0.21  & 7.54  & +0.14 \\
Ca  & 6.36 &  +0.05   & 6.16   &    +0.26 \\
Ti & 5.02  &  -0.01   & 4.98 &    +0.19 \\
\noalign{\smallskip} \hline \end{tabular}
\end{flushleft} 
\end{table*}

\subsubsection{Comparison with Origlia et al. (2002)}

Origlia et al. (2002) analysed 2 stars of 
 T$_{\rm eff}$=4000 K, 
log g=1.0, v$_{\rm t}$=2 km s$^{-1}$,
and a mean metallicity of [FeI/H]=$-$0.30 was derived,
together with enhancement of oxygen [O/Fe]=+0.30,
and [$\alpha$/Fe]=+0.30 (Mg, Ca and Si).
A reddening of E(J-K)=0.41, compatible with E(B-V)=0.70, was used.

\subsubsection{Comparison with Mel\'endez et al. (2003)}

The parameters of II-85, again in common with the present work,
of T$_{\rm eff}$=4000 K, 
log g=1.2, v$_{\rm t}$=1.4 km s$^{-1}$,
[Fe/H]=$-$0.20, found in Mel\'endez et al. (2003),
show an effective temperature higher by 200 K, in agreement with
Barbuy et al. (1999). 
As in Barbuy et al. (1999), the VI colours from HST (Ortolani 
et al. 1995) were used, as well as a reddening
of E(B-V)=0.70.
 Despite the difference with
the present work, the resulting metallicities agree,
which is explained by the low sensitivity of the high excitation
potential infrared FeI lines to T$_{\rm eff}$ (see Table 6 in
Mel\'endez et al. 2003).

\section{Abundance ratios}

\subsection {$\alpha$-elements Mg, Ca, Si, Ti}

The $\alpha$-elements provide information on the 
relative contributions of SNe type II and
Ia in the enrichment of the interstellar medium
prior to the formation of cluster stars. 
 Type II Supernovae are the main sources of $\alpha$-elements
(Woosley \& Weaver 1995), the yields depending upon the mass of the
progenitor.   

\begin{figure}[ht]
\psfig{file=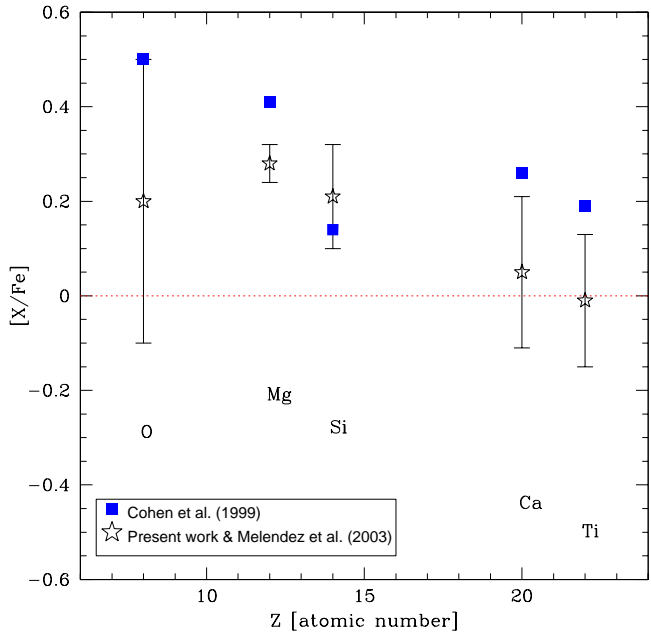,angle=0.,width=9.0 cm}
\caption{Abundance pattern in $\alpha$-elements in NGC 6553 corresponding to:
 (i) a mean of the four sample stars derived in the present work 
and adopting oxygen from Mel\'endez et al. (2003) ({\it open stars})
and (ii) abundances of the same elements given in Table 5 of Cohen
et al. (1999) ({\it filled squares}).}
\label{comp:1}
\end{figure}

Magnesium  and silicon are moderately  enhanced with [Mg/Fe] = +0.28 $\pm$ 0.04
and [Si/Fe] = +0.21 $\pm$ 0.05 dex. A solar calcium-to-iron  [Ca/Fe] = +0.05
$\pm$ 0.09 dex 
 and titanium-to-iron  [TiI/Fe] = $-$0.01 $\pm$ 0.14 and [TiII/Fe]
= $-$0.02 $\pm$ 0.06 are found. 
In Fig. \ref{comp:1} $\alpha$-element abundances from the present work (Table 4),
Mel\'endez et al. (2003), and Cohen et al. (1999) are shown as a function of
Atomic Number Z.
While in Cohen et al. (1999), O, Mg, Si, Ca, and Ti are enhanced, in the present work
Ca and Ti are solar, whereas  Si and Mg are enhanced. 
The high oxygen abundance in Cohen et al. (1999), relative to the value
found by Mel\'endez et al. (2003), may be due to the use of
the OI 771-7 nm triplet that systematically gives higher abundances (Kiselman 1995).
Besides, there is an offset in the solar oxygen abundance adopted (Table 6) of
$\Delta$[O/Fe]=0.09.
 The mean abundance of
[Ti/Fe] in NGC 6553 resembles the value found in Zoccali et al. (2004) 
for NGC 6528, obtained
using the same atomic parameters.

\subsection {Odd-Z elements Na, Al}

Sodium and aluminum are odd-Z elements produced by neutron capture in the Ne-Na
and Mg-Al cycles during carbon burning.
A mean value of [Na/Fe] = +0.16 $\pm$ 0.23 dex and [Al/Fe] = +0.18
$\pm$ 0.06 dex are found. No corrections for non-{\tt LTE} effects were applied. 
 The red giant II-85 of our 
sample is enhanced in Na by [Na/Fe]=+0.34 dex, while the other stars 
show solar ratios. This result is compatible
with a mixing occurring along the RGB (Gratton et al. 2004),
and to check this, it would
be interesting to derive nitrogen abundances for our sample stars.
Na is compatible with results by McWilliam \& Rich 1994, hereafter MR94); 
on the other hand, it is overabundant in  
NGC 6528 (Zoccali et al. 2004). 
A moderate enhancement of Aluminum of [Al/Fe]=+0.18 is found.


\subsection {Heavy elements Zr, Ba, La, Eu}

The s-process elements Zr, Ba, and La are produced through neutron-capture
in asymptotic giant branch stars (K\"appeler 1989; Busso et al. 1999).
Abundances of the s- (Zr, Ba,
and La)  elements obtained were
 [Zr/Fe] = $-$0.67 $\pm$ 0.05, [Ba/Fe] = $-$0.28 $\pm$ 0.18, and
 [La/Fe] = $-$0.11 $\pm$ 0.16. 
Barium is depleted  by $-$0.28 dex in NGC 6553, in contrast to
an enhanced Ba abundance  in 47 Tuc (Alves-Brito et al. 2005), 
a moderate enhancement in MR94, and a solar value in NGC 6528 (Zoccali et
al. 2004). All s-elements are depleted.
Edvardsson et al. (1993) show that
[Ba/H] is a better age indicator than [Fe/H]. 
Low s-element abundances suggest that
the cluster is old, confirming the findings from CMDs (Ortolani et al.
1995). 

The r-element europium abundance is [Eu/Fe] = +0.10 $\pm$ 0.02 dex.
 This moderate overabundance of Eu-to-Fe is compatible with
those of the $\alpha$-elements, these elements being all produced
by SNe II.



\section{Conclusions}

A detailed analysis based on high-resolution spectra for 
four giants of the metal-rich bulge globular cluster NGC 6553
was carried out. The present spectra are the highest resolution data so far for
stars in this cluster.
We find a mean  metallicity value of [Fe/H]=$-$0.20, in agreement with 
Cohen et al. (1999) and Mel\'endez et al. (2003).

 The $\alpha$-elements Mg and Si are
overabundant relative to Fe with [Mg/Fe] = +0.28 dex and [Si/Fe] = +0.21 dex,
which might indicate a rapid chemical evolution history
dominated by Type II Supernovae in the Galactic bulge. 
The lower abundances for
the $\alpha-$elements Ca ([Ca/Fe] = +0.05 dex) and Ti
([Ti/Fe] = $-$0.01 dex) in stars of NGC 6553 might suggest a deficiency of 
low-mass Type II supernovae.
The s-process elements Zr, Ba, and La are depleted by [Zr/Fe]=$-$0.67 dex,
[Ba/Fe]=$-$0.28, and [La/Fe]=$-0.11$ dex, respectively. The
underabundance of s-elements is an independent  indication that the cluster is
old (Edvardsson et al. 1993). The r-process element Eu shows a small
enhancement of [Eu/Fe] =+0.10 dex, as do the $\alpha-$elements Ca and Ti.  
NGC 6553 appears to be very similar in its element abundance ratios to NGC 6528
analysed in Zoccali et al. (2004). The two clusters are very similar in
age and metallicity, as pointed out in Ortolani et al. (1995).

Whether the metal-rich bulge clusters were formed  
during an early collapse or in merging events, as expected 
in hierarchical scenarios, can be revealed by abundance
ratios. An early formation is accompanied by $\alpha$-enhancement.
However in merging events, no detailed models are available:
it is even possible that merging truncates star formation
(Di Matteo et al. 2005).
If instead  bulge formation occurred via secular
evolution of a bar (Kormendy \& Kennicutt 2004),
solar abundance ratios would be expected, but this is not
the case for NGC 6553.
At the present time, it is important to gather metallicities and abundance
ratios for other metal-rich
clusters in the Galactic bulge, since their age spread and chemical abundances
can constrain different models of bulge formation (Bekki 2005; Beasley et al. 2002; 
van den Bergh 1993).

\begin{acknowledgements}

AA acknowledges a Fapesp fellowship, no. 04/00287-9.
BB and EB acknowledges grants from CNPq and Fapesp. 
DM and MZ acknowledge support from the FONDAP Centre for Astrophysics 15010003. 
This work benefited from the
Latin American-European Network on Astrophysics and Cosmology
(LENAC) of the European Union's ALFA Programme.
This publication makes use of data products from the Two Micron All Sky
Survey, which is a joint project of the University of Massachusetts and the
Infrared Processing and Analysis Caltech, funded by the NASA and the NSF.

\end{acknowledgements}


\appendix

\section{Line list}

\begin{longtable}{lccccccccr}

\caption{\label{linelist} Line list used for synthesis showing species,
wavelength, excitation potential, damping constant, gf-values, 
and  abundances derived line-by-line.} \\
\hline\hline

\noalign{\vskip 0.1cm}
\hbox{} & \hbox{} & \hbox{} & \hbox{} & \hbox{} & \multicolumn{4}{c}{[X/Fe]} \\
\cline{6-9} \\

\hbox{Species} & \hbox{$\lambda$({\rm \AA})} & \hbox{$\chi_{ex}$(eV)} &
\hbox{$C6$} & \hbox{log $gf$} &  \hbox{II-64} 
& \hbox{II-85} & \hbox{III-8} & \hbox{267092} \\

\hline
\endfirsthead
\caption{continued.}\\
\hline\hline
\noalign{\vskip 0.1cm}
\hbox{} & \hbox{} & \hbox{} & \hbox{} & \hbox{} & \multicolumn{4}{c}{[X/Fe]} \\
\cline{6-9} \\

\hbox{Species} & \hbox{$\lambda$({\rm \AA})} & \hbox{$\chi_{ex}$(eV)} &
\hbox{$C6$} & \hbox{log $gf$} &  \hbox{II-64} 
& \hbox{II-85} & \hbox{III-8} & \hbox{267092} \\

\hline
\endhead
\hline
\endfoot
\multicolumn{9}{c}{\hbox{\bf $\alpha$-elements}}  \\ 
\noalign{\vskip 0.1cm}
\hbox{Mg I}  & 6318.720 & 5.110 & 0.30E$-$31 & $-2.100$  & +0.27 & +0.27 & +0.27 & +0.12 \\
\hbox{Mg I}  & 6319.242 & 5.110 & 0.30E$-$31 & $-2.360$  & +0.27 & +0.27 & +0.27 & +0.27 \\
\hbox{Mg I}  & 6319.490 & 5.110 & 0.30E$-$31 & $-2.900$  & +0.42 & +0.27 & +0.35 & +0.27 \\

\hbox{Si I}  & 5948.548 & 5.082 & 2.19E-30 & $-1.170$    & +0.20 & +0.20 &+0.25 & +0.15 \\
\hbox{Si I}  & 6142.490 & 5.620 & 0.30E$-$31 & $-1.580$  & +0.20 & +0.12 &+0.10 & +0.20 \\
\hbox{Si I}  & 6145.020 & 5.610 & 0.30E$-$31 & $-1.500$  & ... & +0.25   & ...  & ...	\\ 
\hbox{Si I}  & 6155.020 & 5.620 & 0.30E$-$30 & $-0.850$  & +0.40 & +0.30 &+0.25 & +0.20 \\
\hbox{Si I}  & 6243.823 & 5.610 & 0.30E$-$32 & $-1.300$  & +0.35 & ...   &+0.10 & +0.15 \\
\hbox{Si I}  & 6244.480 & 5.610 & 0.30E$-$31 & $-1.270$  & +0.30 & ...   &+0.15 & +0.15 \\
\hbox{Si I}  & 6414.987 & 5.870 & 0.30E$-$30 & $-1.128$  & +0.20 & ...   &+0.15 & +0.15 \\

\hbox{Ca I}  & 6102.723 & 1.879 & 4.54E$-$31 & $-0.930$  & +0.09 & +0.14 & +0.09 &$-$0.11  \\
\hbox{Ca I}  & 6156.030 & 2.521 & 4.00E$-$31 & $-2.590$  & +0.14 & +0.09 & +0.09 &$-$0.01 \\
\hbox{Ca I}  & 6161.297 & 2.523 & 4.00E$-$31 & $-1.420$  & +0.14 & +0.24 & +0.14 &$-$0.01 \\
\hbox{Ca I}  & 6162.173 & 2.523 & 3.00E$-$31 & $-0.090$  & +0.09 & +0.09 & +0.14 &$-$0.01 \\
\hbox{Ca I}  & 6166.439 & 2.521 & 3.97E$-$31 & $-1.156$  & +0.09 & +0.09 &$-$0.01&$-$0.06 \\
\hbox{Ca I}  & 6169.042 & 2.523 & 3.97E$-$31 & $-0.900$  & +0.29 & +0.19 & +0.14 &$-$0.06 \\
\hbox{Ca I}  & 6169.563 & 2.526 & 4.00E$-$31 & $-0.630$  & +0.29 & +0.19 & +0.14 &$-$0.11 \\
\hbox{Ca I}  & 6439.075 & 2.526 & 3.40E$-$32 & $+0.300$  & +0.14 & +0.09 & +0.14 &$-$0.11 \\
\hbox{Ca I}  & 6455.598 & 2.523 & 3.39E$-$32 & $-1.550$  & +0.19 & +0.09 & +0.14 &  +0.04  \\
\hbox{Ca I}  & 6464.679 & 2.520 & 3.40E$-$32 & $-2.480$  & +0.24 &$-$0.06& +0.24 &  +0.24  \\
\hbox{Ca I}  & 6471.662 & 2.526 & 3.39E$-$32 & $-0.800$  & +0.09 & +0.19 & +0.09 &$-$0.21 \\
\hbox{Ca I}  & 6493.788 & 2.526 & 3.37E$-$32 & $+0.000$  &$-$0.11& +0.19 &$-$0.21&$-$0.21 \\
\hbox{Ca I}  & 6499.654 & 2.526 & 3.37E$-$32 & $-0.850$  &$-$0.11& +0.19 &$-$0.21&$-$0.06 \\
\hbox{Ca I}  & 6508.850 & 2.526 & 3.37E$-$32 & $-2.510$  & +0.09 &$-$0.06&$-$0.06&$-$0.11 \\
\hbox{Ca I}  & 6572.778 & 0.000 & 1.75E$-$32 & $-4.320$  &$-$0.01&...	 &$-$0.16&$-$0.31 \\

\hbox{Ti I}  & 5866.452 & 1.067 & 2.16E$-$32 & $-0.840$  & +0.18 & +0.28 & +0.13  & +0.03  \\ 
\hbox{Ti I}  & 5965.828 & 1.879 & 2.14E$-$32 & $-0.409$  & +0.13 & +0.06 & +0.13  &$-$0.12 \\
\hbox{Ti I}  & 5978.543 & 1.873 & 2.14E$-$32 & $-0.496$  & +0.18 & +0.06 & +0.08  &$-$0.12 \\
\hbox{Ti I}  & 6064.629 & 1.046 & 2.06E$-$32 & $-1.944$  & +0.03 & +0.31 &$-$0.17 &$-$0.17 \\
\hbox{Ti I}  & 6091.174 & 2.267 & 3.89E$-$32 & $-0.423$  & +0.03 & +0.26 & +0.03  &$-$0.12 \\
\hbox{Ti I}  & 6126.217 & 1.067 & 2.06E$-$32 & $-1.424$  & +0.03 & +0.31 & +0.03  &$-$0.47 \\
\hbox{Ti I}  & 6258.110	& 1.440 & 4.75E$-$32 & $-0.360$  &$-$0.07& +0.11 &$-$0.17 &$-$0.52 \\
\hbox{Ti I}  & 6303.756 & 1.443 & 1.53E$-$32 & $-1.566$  & +0.08 & +0.06 &$-$0.07 &$-$0.12 \\
\hbox{Ti I}  & 6312.237 & 1.460 & 4.75E$-$32 & $-1.552$  & +0.08 & +0.11 &$-$0.07 &$-$0.07 \\
\hbox{Ti I}  & 6336.102 & 1.443 & 0.30E$-$31 & $-1.742$  &$-$0.02& +0.11 &$-$0.07 &$-$0.07 \\
\hbox{Ti I}  & 6508.153 & 1.430 & 1.46E$-$32 & $-2.050$  & +0.03 & +0.11 &$-$0.07 &$-$0.07 \\
\hbox{Ti I}  & 6554.224 & 1.443 & 2.72E$-$32 & $-1.219$  & +0.03 & +0.11 &$-$0.17 &$-$0.17 \\
\hbox{Ti I}  & 6556.062 & 1.460 & 2.74E$-$32 & $-1.075$  & +0.03 &$-$0.04&$-$0.17 &$-$0.22 \\
\hbox{Ti I}  & 6599.130 & 0.900 & 2.94E$-$32 & $-2.085$  &$-$0.02& +0.26 &$-$0.07 &$-$0.27 \\
\hbox{Ti I}  & 6743.124 & 0.900 & 0.30E$-$31 & $-1.628$  &$-$0.07& +0.26 &$-$0.07 &$-$0.07 \\
\hbox{Ti II} & 6491.582 & 2.060 & 0.30E$-$31 & $-2.100$  & +0.03 & +0.08  &$-$0.07  &$-$0.07  \\
\hbox{Ti II} & 6559.576 & 2.050 & 0.30E$-$31 & $-2.480$  & +0.09 & $-$0.07&$-$0.02  &$-$0.07  \\
							   
\noalign{\vskip 0.1cm}
\multicolumn{9}{c}{\hbox{\bf Odd-Z elements}}   \\ 
\noalign{\vskip 0.1cm}

\hbox{Na I}  & 5682.647 & 2.102 & 0.30E$-$30 & $-$0.700 & ...  &+0.34&    ... & ...   \\
\hbox{Na I}  & 5688.217 & 2.102 & 0.30E$-$30 & $-$0.457 & ...  &+0.34&    ... & ...   \\
\hbox{Na I}  & 6154.225 & 2.102 & 0.90E$-$31 & $-1.570$ & +0.12&+0.34&$-$0.08 & +0.12 \\
\hbox{Na I}  & 6160.753 & 2.104 & 0.30E$-$31 & $-1.270$ & +0.12& ... &$-$0.08 & +0.12 \\
\hbox{Al I}  & 6696.032 & 3.143 & 0.30E$-$31 & $-1.481$ & +0.28& ... & +0.28  & +0.13 \\
\hbox{Al I}  & 6698.667 & 3.143 & 0.30E$-$31 & $-1.782$ & +0.23& ... & +0.03  & +0.13 \\
							   
\noalign{\vskip 0.1cm}
\multicolumn{9}{c}{\hbox{\bf Heavy elements}}   \\ 
\noalign{\vskip 0.1cm}
\hbox{Zr I}  & 6143.216 & 0.070 & 0.30E$-$31 & $-1.100$  &$-0.74$ & $-0.64$ & $-0.64$ & $-0.64$  \\
\hbox{Ba II} & 6141.728 & 0.704 &  & $-0.070$  &$-0.08$ & $-0.28$ & $-0.58$ & $-0.38$  \\
\hbox{Ba II} & 6496.900 & 0.604 &  & $-0.380$  &$-0.08$ & $-0.08$ & $-0.38$ & $-0.38$  \\
\hbox{La II} & 6390.483 & 0.320 &  & $-1.410$  & +0.01  & $-0.22$ & ...     & ...      \\
\hbox{Eu II} & 6645.127 & 1.380 &  & $+0.120$  & +0.09  &+0.09  & +0.10     & +0.14    \\
\end{longtable}


\begin{longtable}{lcccccccc}

\caption{\label{ironlinelist} Fe I and Fe II line list: (1) element, (2)
wavelength, (3) lower excitation potential, (4) log $gf$-values, (5)-(8)
equivalent widths [m$\rm\AA$].} \\
\hline\hline

\noalign{\vskip 0.1cm}

\hbox{Ion} & \hbox{$\lambda$[{\rm \AA}]} & \hbox{$\chi_{ex}$ [eV]} &
\hbox{log $gf$} &  \hbox{II-64} & \hbox{II-85} & \hbox{III-8} &
\hbox{267092} \\
\hbox{(1)} & \hbox{(2)} & \hbox{(3)} &
\hbox{(4)} &  \hbox{(5)} & \hbox{(6)} & \hbox{(7)} & \hbox{(8)} \\

\hline
\endfirsthead
\caption{continued.}\\
\hline\hline
\noalign{\vskip 0.1cm}

\hbox{Ion} & \hbox{$\lambda$[{\rm \AA}]} & \hbox{$\chi_{ex}$ [eV]} &
\hbox{log $gf$} &  \hbox{II-64} & \hbox{II-85} & \hbox{III-8} &
\hbox{267092} \\
\hbox{(1)} & \hbox{(2)} & \hbox{(3)} &
\hbox{(4)} &  \hbox{(5)} & \hbox{(6)} & \hbox{(7)} & \hbox{(8)} \\
\hline
\endhead
\hline
\endfoot

\noalign{\vskip 0.1cm}
\multicolumn{8}{c}{\hbox{\bf Fe I}}   \\
\noalign{\vskip 0.1cm}

\hbox{Fe I }  & 5833.93 & 2.61  & $-3.66 $   &    ...  &     ...   &   55.2    &   ...    \\
\hbox{Fe I }  & 5835.10 & 4.26  & $-2.37 $   &   39.8  &    30.0   &    ...    &   ...    \\
\hbox{Fe I }  & 5853.15 & 1.48  & $-5.27 $   &   59.7  &    88.3   &   44.8    &   48.0   \\
\hbox{Fe I }  & 5855.08 & 4.60  & $-1.75 $   &   39.6  &    47.6   &   41.6    &   38.7   \\
\hbox{Fe I }  & 5856.09 & 4.29  & $-1.64 $   &   61.7  &    66.7   &   57.3    &   62.0   \\
\hbox{Fe I }  & 5858.78 & 4.22  & $-2.26 $   &   29.3  &    36.6   &   28.8    &   28.9   \\
\hbox{Fe I }  & 5859.60 & 4.55  & $-0.60 $   &   94.7  &    88.5   &   91.9    &   91.1   \\
\hbox{Fe I }  & 5861.11 & 4.28  & $-2.45 $   &   20.6  &    24.9   &   16.9    &   26.5   \\
\hbox{Fe I }  & 5862.37 & 4.55  & $-0.39 $   &  108.1  &    95.9   &  102.6    &  107.1   \\
\hbox{Fe I }  & 5881.28 & 4.60  & $-1.84 $   &    ...  &    34.8   &   38.0    &   35.1   \\
\hbox{Fe I }  & 5902.47 & 4.59  & $-1.81 $   &   24.1  &    24.5   &   26.2 
&   23.8   \\
\hbox{Fe I }  & 5905.67 & 4.65  & $-0.72 $   &   74.7  &    67.4   &   78.3    &   74.4   \\
\hbox{Fe I }  & 5927.79 & 4.65  & $-1.09 $   &   64.7  &    65.1   &   67.8    &   65.6   \\
\hbox{Fe I }  & 5929.68 & 4.55  & $-1.39 $   &   75.2  &    63.8   &   75.6    &   61.5   \\
\hbox{Fe I }  & 5930.18 & 4.65  & $-0.23 $   &  121.3  &   102.6   &  118.7    &  120.0   \\
\hbox{Fe I }  & 5934.65 & 3.93  & $-1.18 $   &  111.3  &   106.0   &  106.9    &  108.6   \\
\hbox{Fe I }  & 5952.72 & 3.98  & $-1.43 $   &   97.2  &    82.0   &   91.0    &   94.7   \\
\hbox{Fe I }  & 5956.69 & 0.86  & $-4.60 $   &  137.6  &     ...   &  126.1    &  120.6   \\
\hbox{Fe I }  & 5969.56 & 4.28  & $-2.73 $   &    ...  &    15.8   &    ...    &    ...   \\
\hbox{Fe I }  & 5983.69 & 4.55  & $-0.78 $   &   97.5  &    83.4   &   81.8    &   91.6   \\
\hbox{Fe I }  & 5987.07 & 4.79  & $-0.45 $   &   93.2  &    84.7   &   92.0    &   95.9   \\
\hbox{Fe I }  & 6003.01 & 3.88  & $-1.11 $   &  119.7  &   106.9   &  119.2    &  113.9   \\
\hbox{Fe I }  & 6024.06 & 4.55  & $-0.11 $   &  124.8  &   107.7   &  121.0    &  119.1   \\
\hbox{Fe I }  & 6027.05 & 4.07  & $-1.22 $   &   98.4  &    87.5   &   79.1    &   95.5   \\
\hbox{Fe I }  & 6054.07 & 4.37  & $-2.30 $   &   22.2  &    30.3   &   27.1    &   23.6   \\
\hbox{Fe I }  & 6056.01 & 4.73  & $-0.46 $   &   85.5  &    83.4   &  112.0    &   90.8   \\
\hbox{Fe I }  & 6078.50 & 4.79  & $-0.40 $   &   99.0  &    86.9   &   96.2    &   95.4   \\
\hbox{Fe I }  & 6079.01 & 4.65  & $-1.13 $   &   67.8  &    64.5   &   66.3    &   68.6   \\
\hbox{Fe I }  & 6082.71 & 2.22  & $-3.57 $   &  100.9  &   131.1   &   87.5    &   90.0   \\
\hbox{Fe I }  & 6093.64 & 4.60  & $-1.51 $   &   53.0  &    49.8   &   53.0    &   48.8   \\
\hbox{Fe I }  & 6094.37 & 4.65  & $-1.94 $   &   37.4  &    35.9   &   35.2    &   34.1   \\
\hbox{Fe I }  & 6096.66 & 3.98  & $-1.93 $   &   68.5  &    70.1   &   67.2    &   66.1   \\
\hbox{Fe I }  & 6098.28 & 4.56  & $-1.88 $   &   33.0  &    46.5   &   39.1    &   37.0   \\
\hbox{Fe I }  & 6105.13 & 4.54  & $-2.06 $   &   31.8  &    43.0   &   30.2    &   28.0   \\
\hbox{Fe I }  & 6120.25 & 0.91  & $-5.95 $   &   59.7  &    96.4   &   52.4    &   45.0   \\
\hbox{Fe I }  & 6353.85 & 0.91  & $-6.36 $   &    ...  &    71.7   &    ...    &    ...   \\
\hbox{Fe I }  & 6392.54 & 2.28  & $-4.03 $   &   71.7  &     ...   &   56.9    &   57.0   \\
\hbox{Fe I }  & 6419.95 & 4.73  & $-0.25 $   &  107.5  &    97.8   &  101.5    &  105.1   \\
\hbox{Fe I }  & 6475.62 & 2.56  & $-2.94 $   &  121.3  &   127.5   &  110.5    &  103.3   \\
\hbox{Fe I }  & 6481.87 & 2.28  & $-2.98 $   &  141.2  &   148.5   &  121.1    &  123.1   \\
\hbox{Fe I }  & 6498.94 & 0.96  & $-4.70 $   &  140.4  &     ...   &  124.6    &  126.6   \\
\hbox{Fe I }  & 6518.37 & 2.83  & $-2.75 $   &  105.0  &   120.1   &  102.8    &  105.8   \\
\hbox{Fe I }  & 6533.93 & 4.56  & $-1.45 $   &   69.5  &    58.5   &   61.3    &   66.8   \\
\hbox{Fe I }  & 6556.81 & 4.79  & $-1.68 $   &   24.3  &    28.4   &   30.0    &   25.6   \\
\hbox{Fe I }  & 6569.22 & 4.73  & $-0.42 $   &  105.5  &   102.0   &  100.2    &  101.6   \\
\hbox{Fe I }  & 6574.23 & 0.99  & $-5.04 $   &  114.3  &   149.3   &  102.3    &   98.5   \\
\hbox{Fe I }  & 6575.02 & 2.59  & $-2.82 $   &    ...  &     ...   &  119.3    &  120.3   \\
\hbox{Fe I }  & 6591.31 & 4.59  & $-2.06 $   &   23.4  &     ...   &   21.6    &   25.6   \\
\hbox{Fe I }  & 6593.87 & 2.43  & $-2.42 $   &    ...  &     ...   &  142.2    &  148.5   \\
\hbox{Fe I }  & 6597.56 & 4.80  & $-1.06 $   &   64.4  &    60.3   &   62.1    &   69.6   \\
\hbox{Fe I }  & 6608.03 & 2.28  & $-4.04 $   &   70.5  &     ...   &   59.7    &   56.6   \\
\hbox{Fe I }  & 6699.14 & 4.59  & $-2.18 $   &   29.4  &    25.8   &   24.7    &   20.7   \\
\hbox{Fe I }  & 6703.57 & 2.76  & $-3.15 $   &   93.4  &    96.5   &   82.6    &   84.5   \\
\hbox{Fe I }  & 6705.11 & 4.61  & $-1.40 $   &   77.2  &    58.5   &   72.7    &   74.2   \\
\hbox{Fe I }  & 6710.32 & 1.48  & $-4.87 $   &   94.6  &   112.1   &   83.1    &   79.3   \\
\hbox{Fe I }  & 6713.74 & 4.80  & $-1.60 $   &   40.0  &    33.1   &   36.7    &   37.9   \\
\hbox{Fe I }  & 6715.38 & 4.59  & $-1.64 $   &   65.3  &     ...   &   57.6    &   ...    \\
\hbox{Fe I }  & 6725.36 & 4.10  & $-2.30 $   &   41.1  &    41.5   &   39.2    &   41.5   \\
\hbox{Fe I }  & 6726.67 & 4.59  & $-1.15 $   &   71.5  &    56.4   &   67.6    &   66.7   \\
\hbox{Fe I }  & 6733.15 & 4.64  & $-1.58 $   &   48.7  &    54.6   &   52.8    &   50.4   \\
\hbox{Fe I }  & 6739.52 & 1.56  & $-4.94 $   &   72.2  &    98.4   &   64.0    &   57.2   \\
\hbox{Fe I }  & 6752.71 & 4.64  & $-1.37 $   &   74.7  &     ...   &   66.4    &   62.8   \\
\noalign{\vskip 0.1cm}
\noalign{\vskip 0.1cm}\multicolumn{8}{c}{\hbox{\bf Fe II}}     \\

\hbox{Fe II}  & 5991.36 & 3.15  &  -3.54   & ...  &  ... & 43.4 &  ... \\
\hbox{Fe II}  & 6084.09 & 3.20  &  -3.79   & 31.2 &  ... & 31.7 &  ... \\
\hbox{Fe II}  & 6149.24 & 3.89  &  -2.69   & ...  &  ... & 51.4 &  ... \\
\hbox{Fe II}  & 6247.56 & 3.89  &  -2.30   & ...  &  ... & 50.7 & 54.7  \\
\hbox{Fe II}  & 6432.68 & 2.89  &  -3.57   & 42.0 & 29.1 & 53.8 & 55.9  \\
\hbox{Fe II}  & 6456.39 & 3.90  &  -2.05   & 69.3 & 33.5 & 72.2 & 64.5  \\
\hbox{Fe II}  & 6516.08 & 2.89  &  -3.31   & 66.9 &  ... &  ... & 69.3  \\

\end{longtable}

\end{document}